\documentclass[aps,prb,twocolumn,superscriptaddress]{revtex4}

\usepackage{graphicx}
\usepackage{dcolumn}
\usepackage{bm}
\usepackage[normalem]{ulem}

\begin{document}

\title{Phase diagram of H$_{\bm 2}$ adsorbed on graphene
}  

\author{M.C. Gordillo}

\affiliation{Departamento de Sistemas F\'{\i}sicos, Qu\'{\i}micos 
y Naturales, Facultad de Ciencias Experimentales, Universidad Pablo de
Olavide, Carretera de Utrera, km 1. 41013 Sevilla, Spain}

\author{J. Boronat}

\affiliation{Departament de F\'{\i}sica i Enginyeria Nuclear, 
Universitat Polit\`ecnica de Catalunya, 
Campus Nord B4-B5, 08034 Barcelona, Spain}

\date{\today}

\begin{abstract}
The phase diagram of the first layer of H$_2$ adsorbed on top of a single graphene
sheet has been  calculated by means of a series of diffusion Monte
Carlo (DMC) simulations. We have found that, as in the case of $^4$He,
the ground state of molecular hydrogen is a
$\sqrt3 \times \sqrt3$ commensurate structure, followed, upon a pressure
increase, by an incommensurate triangular solid. A striped phase 
of intermediate density was also considered, and found lying on top of the 
equilibrium curve separating both commensurate and incommensurate solids.   
\end{abstract}


\maketitle

\section{Introduction}

Graphene is a novel form of carbon, in which the atoms 
are located in the nodes of a honeycomb lattice that 
extends periodically in two dimensions forming a single
layer.~\cite{science2004,pnas2005} In this respect, graphene is
different from graphite, a substrate formed by the superposition
of these layers to build a complete three dimensional structure. 
Graphene can be obtained
as a free standing structure,~\cite{pnas2005} or as a single adsorbed layer
on top of another substrate.~\cite{nature2009} 
One of the main interests of the scientific community on graphene
is related to its unusual electron transport properties 
that are determined by the Dirac equation,~\cite{nature1,nature2,jpcbgra,natmat} 
but other characteristics of this compound
as for instance its behavior as adsorber are starting to be payed attention
to. A recent diffusion Monte Carlo (DMC) calculation of $^4$He on graphene 
indicates that its behavior is quite similar to the one  
on graphite, the main difference being the binding energy, lower
in the case of a single carbon sheet.~\cite{yo} It was also
found~\cite{yo} that the ground state of helium on graphene is a $\sqrt3 \times
\sqrt3$ commensurate solid, in agreement with experimental data on 
graphite.~\cite{grey,grey2} 
The aim of the present work is to calculate the phase diagram of  H$_2$
adsorbed on graphene at zero temperature using for the first time the
DMC method. The comparison with the case of graphite will show us if 
there is any significant difference between them or graphene is simply
a weaker binding version of the graphite phase diagram, as happens with
$^4$He.

\begin{table*}
\caption{Equilibrium density $\rho_0$ and energy per H$_2$ molecule at equilibrium $e_0$ 
for the  liquid phase. Graphene (2D) and graphite (2D) indicate the results
after the subtraction of their respective energies in the infinite
dilution limit. 2D are DMC results for a strictly 2D H$_2$
system with the same intermolecular potential (Ref. \onlinecite{claudio}). 
The spinodal densities $\rho_{\text s}$ for the same systems are also shown. 
}
\begin{tabular}{cccccc} \hline
       & Graphene &  Graphite & Graphene (2D) & Graphite (2D) & 2D \\  \hline
$e_{0} (K)$ & $-451.88 \pm  0.03$ & $-503.0 \pm  0.1$ & $-20.09 \pm  0.07$
&   $-20.4 \pm  0.1$ & $-21.43  \pm 0.02$ \\
$\rho_{0}$ (\AA$^{-2}$) &  $0.05948 \pm 0.00005$ & $0.0593 \pm 0.0004$ &
$0.05948 \pm 0.00005$ & $0.0593 \pm 0.0004$ & $0.0633 \pm 0.0003$ \\ 
$\rho_{\text s}$ (\AA$^{-2}$) &  $0.0489 \pm 0.0001$ & $0.0486 \pm 0.0001$
& $0.0489 \pm 0.0001$ & $0.0486 \pm 0.0001$ & - \\ 

 \hline
\end{tabular}
\end{table*}

\section{Method}

Our study is based on the DMC method, basically because this 
technique allow us to obtain the correct ground state for a given  
system of bosons.~\cite{boro94}  This is the case here, since we will consider  
only para-H$_2$, the ground state of the hydrogen molecule.  
An important ingredient of any DMC calculation is the trial wave function
used for importance sampling. This function collects 
basic information about the system that is known {\em a priori}, and can be considered
a reasonable approximation (in the variational sense) to its ground state.
In this work, we use as a intial trial wave function  
\begin{eqnarray}
\lefteqn{ \Phi({\bf r}_1, {\bf r}_2, \ldots, {\bf r}_N)  =  \prod_{i<j} \exp \left[-\frac{1}{2} 
\left( \frac{b_{{\text H}_2{\text -}{\text H}_2}}{r_{ij}} \right)^5
\right]}  \label{trial1} \\ 
& & \times \prod_i  \prod_J \exp \left[ -\frac{1}{2} \left( \frac{b_{{\text
C}{\text -}{\text
H}_2}}{r_{iJ}} \right)^5 \right] 
\prod_i \exp (-a (z_i-z_0)^2) \nonumber \ ,
\end{eqnarray} 
that depends on the coordinates of the hydrogen molecules
${\bf r}_1$, ${\bf r}_2$, \ldots, ${\bf r}_N$ and the position of the
Carbon atoms ${\bf r}_J$ of the substrate. The first term in Eq.
(\ref{trial1})  is a two-body Jastrow function depending on the
H$_2$ intermolecular distances $r_{ij}$, with optimal parameter 
$b_{{\text H}_2\text{-H}_2} = 3.195$ \AA. 
The second term is another Jastrow function that takes into account all the individual 
C-H$_2$  interactions, the optimal value of $b_{\text{C-H}_2}$ being 2.3 \AA. 
Finally, the third term is a product of one-body Gaussians 
that depend only on the $z$ coordinate
of each molecule and whose function is to localize the molecules near the
$z_0$ value where the binding energy
formed by the summing up of all the carbon-hydrogen interaction in
the graphene-adsorbate system is larger. It depends on two parameters, whose
optimal values are $a = 3.06$ \AA$^{-2}$  and $z_0 = 2.9$ \AA.
All the parameters in the trial wave function (\ref{trial1})  
were variationally optimized for a liquid phase at density  0.0068 \AA$^{-2}$ 
and their slight density dependence was neglected.

In the case of the different solid phases considered in this work, the trial function
above (\ref{trial1}) was multiplied by another set of Gaussians of the form
\begin{equation}
\prod_i \exp\{-c[(x_i-x_{\text{site}})^2+ (y_i - y_{\text{site}})^2]\}
\label{trialsol}
\end{equation}
i.e., each particle was limited to be in a region around its corresponding 
$x_{\text{site}},y_{\text{site}}$ coordinates (Nosanow-Jastrow model). 
These coordinates corresponded to
the crystallographic positions of the particular solid considered, and the
parameter $c$
was variationally optimized for each lattice type. For instance, for
all the commensurate phases considered in this work, $c = 0.61$ \AA$^{-2}$. 
For the incommensurate solid, a linear fit between the values obtained
at densities 0.1 ($c= 1.38$ \AA$^{-2}$) and 0.08 \AA$^{-2}$ ($c = 0.61$ \AA$^{-2}$) was used. 
Below the latter density, $c$ was kept fixed to 0.61 \AA$^{-2}$.  
We used the Silvera and Goldman potential~\cite{silvera} for the
H$_2$-H$_2$ interaction, one of the  
standard potentials in Monte Carlo calculations of para-H$_2$. 
It considers point-like H$_2$ molecules, due to the low excentricity of
the ellipsoid of the real molecule, and it has been shown that reproduces
quite accurately  the equation of state of bulk solid H$_2$. 
The corrugation effects induced by the substrate were introduced by 
taking into account all the C-H$_2$ individual interactions, that are
assumed to be of Lennard-Jones type  with parameters taken from Ref. 
\onlinecite{coleh2}.  All the above sets of parameters were used both for graphene
and graphite alike, the only difference in their respective simulations being the number 
of graphene
sheets considered in the calculation of the C-H$_2$ interaction potential.  
Following the results of Ref. \onlinecite{yo}, we modeled graphite by a series
of 8 parallel graphene layers separated by a distance of 3.35 \AA from each other 
and stacked in the A-B-A-B form characteristic of this compound. 
The presence of additional carbon layers did not modify the H$_2$ binding energies 
to the substrate, the results for 8 and 9 carbon sheets being equivalent within 
their error bars.  

\section{Results}

We started our study by comparing the experimentally obtained binding
energies $e_b^{'}$s in the infinite dilution limit to the ones calculated
by means of the DMC technique described above. This was done for graphite,
since there is no experimental data for graphene yet. The results are $e_b =
-482.7 \pm 3.5$ K (experimental, Ref. \onlinecite{colesur}) 
versus $-482.57 \pm 0.06$ K
(this work).  The good agreement between both values means 
that the C-H$_2$ potential used here is expected
to give accurate results for this system. The energy obtained for graphene
was $-431.79 \pm 0.06$ K, i.e., a difference with the previous case of
$50.78 \pm 0.08$ K.

\begin{figure}[b]
\begin{center}
\includegraphics[width=0.8\linewidth]{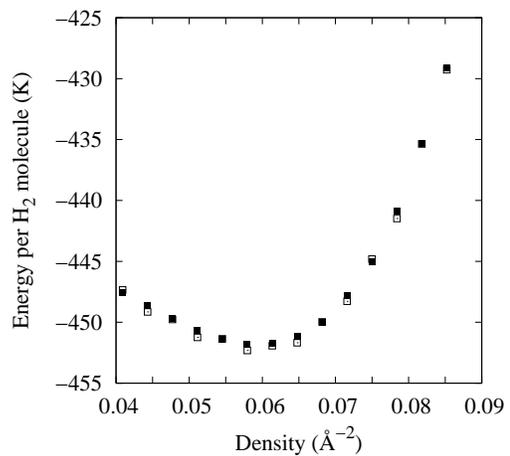}
\caption{Energy per H$_2$ molecule in the metastable liquid phase for
graphene (full boxes) and graphite (open boxes). To afford an easy 
comparison, this latter results are up-shifted by the difference between 
the binding energies of a single hydrogen molecule on top of graphene
and graphite.  
}
\end{center}
\label{fig1}
\end{figure}

\begin{table*}
\caption{Density and energy per H$_2$ molecule for the incommensurate triangular 
solids obtained by least squared fits to the simulation results.  
Graphene (2D) and graphite (2D) indicate the results after the subtraction
of their respective energies in the infinite dilution limit, as in the previous table. 
The 2D entries are results 
for different calculations of pure 2D H$_2$ systems. 
}
\begin{tabular}{ccc ccc cc} \hline
       & Graphene &  Graphite & Graphene (2D) & Graphite (2D) & 2D (Ref. \onlinecite{claudio}) & 2D (Ref. \onlinecite{prl97}) & 2D (Ref. \onlinecite{bonin}) \\  \hline
$e_{0} (K)$ & $-454.1 \pm  0.3$ & $-505.2 \pm  0.2$ & $-22.3 \pm  0.3$ & 
$-22.6 \pm  0.2$ & $-23.453  \pm 0.003$ & $-22.1 \pm 0.1$ & $-23.25 \pm
0.05$ \\
$\rho_{0}$ (\AA$^{-2}$) &  $0.0689 \pm 0.0005$ & $0.0689 \pm 0.0006$ &
$0.0689 \pm 0.0005$ & $0.0689 \pm 0.0006$ & $0.0673 \pm 0.0002$ & $0.064
\pm 0.01$ & $0.0668 \pm 0.0005$ \\ 
$\rho_{\text s}$ (\AA$^{-2}$) &  $0.0606 \pm 0.0001$ & $0.0600 \pm 0.0001$
& $0.0606 \pm 0.0001$ & $0.0606 \pm 0.0001$ & $0.0584 \pm 0.0001$ & - &
$0.059 \pm 0.001$ \\ 
 \hline
\end{tabular}
\end{table*}

\begin{figure}[b]
\begin{center}
\includegraphics[width=0.8\linewidth]{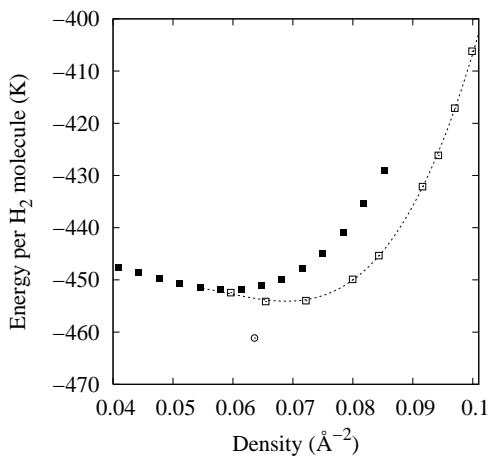}
\caption{Adsorption energy of  H$_2$ on graphene. Full boxes, liquid
metastable phase; open boxes, incommensurate triangular solid; open 
circle, commensurate $\sqrt3 \times \sqrt3$ registered solid.  
}
\end{center}
\label{fig2}
\end{figure}

\begin{figure}[b]
\begin{center}
\includegraphics[width=0.8\linewidth]{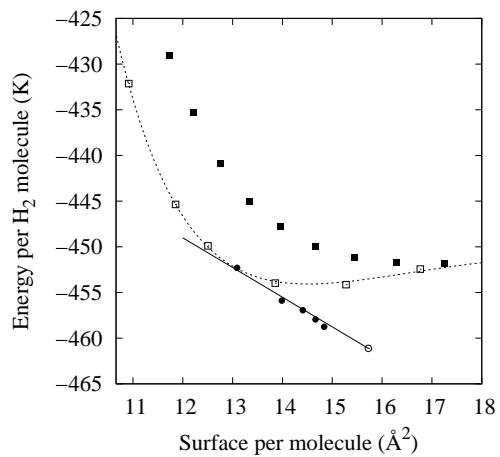}
\caption{Same than in previous figure but including the data 
for the $\alpha$ phase (full circles). The x-axis represents now  
the inverse of the density. The error bars are of the size of the
symbols and not displayed for symplicity.  
}
\end{center}
\label{fig3}
\end{figure}

Our main goal was to calculate the phase diagram of the first layer
of H$_2$ on top of both
graphene and graphite until  the experimental density for promotion to the
second layer, 0.0937 \AA$^{-2}$~\cite{colebook}   and compare the results
obtained. The first step towards this end is reported in Fig. 1, where we
show the energies per particle of a liquid phase on top of graphene 
(black boxes) and graphite (open boxes). 
Even though the liquid phase will be shown to be metastable,  
it is useful to calculate
its properties from a methodological point of view,  
for the sake of comparison with other possible phases and previous calculations. 
In both cases, we used the same simulation cell of
$34.43 \times 34.08$ \AA$^2$, with  a variable number of H$_2$ molecules on
top of it to match the densities displayed in the  figure. Periodic
boundary conditions were considered for the $x$ and $y$ directions. To
aid in the comparison, the results for graphite were up-shifted by the
difference in the binding energies in the infinite dilution limit between
graphene and graphite, given in the last paragraph. Since the error bars
are of the size of the symbols, we can see that both curves are virtually
identical. This is confirmed by a look to the results displayed in Table I,
that come from least-squared fitting with cubic polynomials to
the DMC data displayed in Fig. 1. The equilibrium densities $\rho_0$ 
in graphene and graphite are equal within error bars, 
and the difference between the energies per molecule at equilibrium $e_0$
is  $\sim 0.3$ K after
having considered the shifting due to the differences in the substrate. 
These results indicate that in the metastable liquid phase,  
the main difference
between graphene and graphite as adsorbents is a nearly constant shift
in the binding
energies, as in the case of $^4$He.~\cite{yo} 
The spinodal densities $\rho_{\text s}$, defined as the points in which 
the derivative of the pressure with respect to the surface density equals
zero, are also indistinguishable in both cases. 
 We can compare the results obtained for the liquid phase with
the ones for a purely two dimensional (2D) system. In Ref.
\onlinecite{claudio}, this was made in the same conditions that the ones
considered here,  i.e., with the same intermolecular potential and for
$T=0$ K,
as corresponds to a DMC calculation; their results  are shown in Table I
for comparison. We
can infer that the introduction of both corrugation and movement in a
perpendicular direction to the basal plane have the effect of decreasing
the binding energy of the liquid at zero pressure (21.43 for a flat
surface versus around 20 K for graphene and graphite).

Experimental data of H$_2$ on top of graphite~\cite{colebook,frei1,frei2,frei3} 
indicate that the liquid phase is  metastable
with respect to both the incommensurate and commensurate solids, as in
the $^4$He case.  This feature is confirmed by the present results shown in Fig.
2, where we display the energies of H$_2$ on graphene for its liquid (full boxes),
commensurate $\sqrt3 \times \sqrt3$ structure (open circle, density 0.0636
\AA$^{-2}$), and incommensurate triangular solid phases (open boxes). 
The dotted line is a least-squares fit to DMC data for the latter phase
with a third degree polynomial. 
From  Fig. 2, we conclude  that the  ground state
of H$_2$ on top of graphene is a $\sqrt3 \times \sqrt3$ commensurate (C)
solid, as it is on graphite,  whose phase diagram is similar and not shown
for simplicity. The binding energies  per H$_2$ molecule for
this C solid are $-461.12 \pm 0.01$ and $-512.97 \pm 0.02$ K for graphene
and graphite,  respectively. Their difference (51.85 K) is slightly larger
than the one for the  binding energy in the infinite  dilution
limit (50.78 K), indicating that the additional stabilization due to the
extra graphite layers is more important for the C solid phase 
(1.07 versus 0.37 K). The energy of the C phase for graphite 
is slightly below the variational  results of Novaco~\cite{novaco} for the
same structure ($-510.8$ K). Since the binding energy of this commensurate 
phase is  always
larger than the one corresponding to the liquid phase, we conclude that for
densities smaller than 0.0636 \AA$^{-2}$ the system will break in H$_2$
patches separated by empty space to produce the average density we could be
interested in.

In order to establish the minimum density for which the incommensurate (IC) 
solid is
stable we should make a  double-tangent Maxwell construction between 
the C and IC phases.
Since the C phase is defined by a single density, the construction was made by 
drawing the tangent line to the IC equation of state that intersects the C 
point.
The result is depicted in Fig. 3 for the case of
graphene. There, we can see that the limiting density for the IC solid in
equilibrium with the $\sqrt3 \times \sqrt3$ structure is 0.077
\AA$^{-2}$ (the inverse of a limiting surface per molecule of 13 \AA$^2$), 
in excellent agreement with the available experimental data for
graphite~\cite{frei1,frei2,frei3}  between 9 and 20 K (0.077 $\pm$ 0.001 \AA$^{-2}$ $\sim$ 1.22 times
the density of the C phase). The binding energy corresponding to this 
structure is $-452.08$ K.
The lower density for graphite is exactly the same, with a binding
energy of $-503.15$ K.  Obviously, to draw the limits of this transition
is only  possible if we consider corrugation in the graphene and
graphite structures.  The limits for this C-IC transition were not
calculated quantitatively in any of the previous works of H$_2$ on 
graphite.~\cite{ni,got,novaco,manou1,manou2,vives}  
There have been also other calculations
on the equation of state for a triangular solid in a purely 2D environment,
both at zero~\cite{claudio} and finite temperature.~\cite{prl97,bonin} Their
main results are summarized in Table II and compared to the present results for the 
IC solid. However, in the context of H$_2$ adsorption on graphene and
graphite these strictly 2D results only serve as a check of the quality of our
calculations, since the equilibrium density  of the triangular 2D solid is
below the IC density limit determined by the Maxwell construction (Fig. 3).    

\begin{table}
\caption{Energy per H$_2$ molecule for different arrangements found in the phase diagram 
of D$_2$ on graphite and its corresponding values 
for an incommensurate triangular solid at the same density. 
All data correspond to graphene.}  
\begin{tabular} {cccc} \hline
Phase & $\rho$ (\AA$^{-2}$) & $e_{\text{binding}}$ (K)  & $e_{\text{IC}}$ (K) \\ \hline
$\delta$  &  0.0789 & $-442.22 \pm 0.08$  & $-450.88 \pm 0.1$ \\   
$\epsilon$  & 0.0835 & $-440.30 \pm 0.04$ & $-446.47 \pm 0.1$ \\
$\gamma$  & 0.0814 & $-441.24 \pm 0.06$ & $-448.76 \pm 0.1$ \\ 
\hline
\end{tabular}
\end{table}

We have also analyzed the possible  existence of the so-called $\alpha$
phase, or striped domain phase, in adsorbed H$_2$. According to experiments
in graphite,~\cite{frei2,cui} it consists of strips of the C phase of
variable width separated by narrow walls in which the H$_2$ molecules are
closer to each other (see the phase diagrams in Ref. \onlinecite{frei2}). 
This phase is also present in D$_2$,~\cite{deu3175} and appears in the
phase diagram drawn in Ref \onlinecite{manou2}.  The $\alpha$  structure
can be defined by a rectangular cell that contains more or less H$_2$
molecules depending on the width of the C domains. In Fig. 3, we display 
as  solid circles the cases for 4,6,8,10 and 12 molecules per simulation
cell. It can be seen that they are approximately on top of the Maxwell
construction line that limits the C-IC transition. 
 This result implies
that we can consider the $\alpha$ phase as intermediate between them, with
continuous changes from the C to the $\alpha$ phase, and from the $\alpha$
to the IC solid, at least at $T=0$ K. This continuous change is in
agreement  with the experimental results of Cui {\em et al},~\cite{cui}
which ruled out a first order change between the C and the $\alpha$
arrangement. However, our results do not allow us to  determine if the
$\alpha$ phase is a real thermodynamical phase or simply a mixture  of the
C and IC phases separated by striped domains. The reason is that the
simulation  results are below but too close to the
Maxwell construction line for making a definitive commitment (see Fig. 3).  
The data for
graphite are similarly basicaly on top of the  corresponding Maxwell
construction line, and can be obtained from the graphene ones in Fig. 3 by
applying a downward shift of 51.46 K. 

To complete the study of H$_2$ on top of graphene, we checked the existence
of three hypothetical phases that appear in the experimental~\cite{deu3175}
 phase diagram
of D$_2$  on top of graphite and that have not been
experimentally observed in H$_2$. Two of them 
are commensurate: the $\epsilon$ phase, which is a $4 \times 4$ structure ($\rho = 0.0835$
\AA$^{-2}$), and the $\delta$ phase, which corresponds to a
$5 \sqrt3 \times 5 \sqrt3$ arrangement ($\rho = 0.0789$
\AA$^{-2}$). Both of them are within the  density limits corresponding to
the IC triangular phase. In D$_2$ on graphite, there is also an
incommensurate oblique phase, which is called the $\gamma$ phase, whose
range of stability is approximately between  0.077 and 0.083 \AA$^{-2}$. We
calculated the corresponding energies per particle for the above mentioned
commensurate phases and for a single density of the $\gamma$ phase (0.0814
\AA$^{-2}$), and compared them to the energies of an IC triangular phase at
the same densities. The results are displayed in Table III.
The main conclusion is that for H$_2$ all these arrangements are metastable with
respect to the incommensurate solid. 

\section{Concluding remarks}

Summarizing, we have studied the phase diagram of H$_2$ on top of graphene 
using the most powerful microscopic tool at zero temperature (DMC), 
accurate potentials, and incorporating explicit C-H$_2$ interactions (fully
corrugated model). The phase diagram of the first layer on graphene 
is fully determined for the first time. The ground state corresponds to a
$\sqrt3 \times \sqrt3$ commensurate solid as happens in $^4$He.~\cite{yo}
Graphene and graphite show basically the same phase diagram,  
the main difference being the adsorption
energy,  which is $\sim$ 51 K larger in graphite.   

We acknowledge partial financial support from the 
Junta de Andaluc\'{\i}a group PAI-205, DGI (Spain) Grant No.
 FIS2008-04403 and Generalitat de Catalunya Grant No. 2009SGR-1003.



\begin{thebibliography}{99}

\bibitem{science2004} K.S. Novoselov, A.K. Geim, S.V. Morozov, D. Jiang, 
Y. Zhang, 
S.V. Dubonos, I.V. Grigorieva, and A.A. Firsov, Science {\bf 306},
666 (2004).  

\bibitem{pnas2005} K.S. Novoselov, D. Jiang, F. Schedin, T.J. Booth, 
V.V. Khotkevich, S.V. Morozov, and A.K. Geim, PNAS {\bf 102}, 10451 (2005).

\bibitem{nature2009} C.H. Lui, L. Liu, K.F. Mak, G.W. Flynn, and T.F. Heinz, 
Nature (London) {\bf 462}, 339 (2009). 

\bibitem{nature1} K.S. Novoselov, A.K. Geim, S.V. Morozov, D. Jiang,
M.I. Katsnelson,  I.V. Grigorieva, S.V. Dubonos, and A.A. Firsov, Nature (London)
{\bf 438}, 197 (2005).

\bibitem{nature2} Y. Zhang, Y. Tan, H.L. Stormer, and P. Kim.
Nature (London) {\bf 438}, 201 (2005). 

\bibitem{jpcbgra} C. Berger, Z. Song, T. Li, X. Li, A.Y. Ogbazghi, 
R. Feng, Z. Dai, A.N. Marchenkov, E.H. Conrad, P.N. First, and W.A. Heer, 
J. Phys. Chem B. {\bf 108}, 19912 (2004). 

\bibitem{natmat} A.K. Geim and K.S. Novoselov, Nat.Mat. {\bf 6}, 183 (2007). 

\bibitem{yo} M.C. Gordillo and J. Boronat, Phys. Rev. Lett. {\bf 102},
085303 (2009). 

\bibitem{grey} D.S. Greywall and P.A. Busch, Phys. Rev. Lett. {\bf 67},
3535 (1991).  

\bibitem{grey2} D.S. Greywall, Phys. Rev. B {\bf 47},  309 (1993).

\bibitem{boro94} J. Boronat and J. Casulleras, Phys. Rev. B {\bf 49},
8920 (1994).

\bibitem{silvera} I. F. Silvera and V. V. Goldman, J. Chem. Phys. {\bf 69}, 4209
(1978).

\bibitem{coleh2} G. Stan and M.W. Cole, J. Low Temp. Phys. {\bf 110}, 539 (1998).

\bibitem{colesur}  G. Vidali, G. Ihm, H.Y Kim, and  M.W. Cole, Sur. Sci. Reports 
{\bf 12}, 135 (1991).

\bibitem{colebook} L.W. Bruch, M.W. Cole, and E. Zaremba, \textit{Physical
adsorption: forces and phenomena}, Oxford University Press, Oxford (1997). 

\bibitem{claudio} C. Cazorla and J. Boronat, Phys. Rev. B. {\bf 78}, 134509 (2008).  

\bibitem{frei1} H. Freimuth and H. Wiechert, Surf. Sci. {\bf 162}, 432 (1985).  

\bibitem{frei2} H. Freimuth and H. Wiechert, Surf. Sci. {\bf 189/190}, 548 (1987).  

\bibitem{frei3}  H. Wiechert, Physica B {\bf 169}, 144 (1991).  

\bibitem{cui} J. Ciu and S.C. Fain, Jr. Phys. Rev. B {\bf 39} 8628 (1989).  

\bibitem{novaco} A.D. Novaco, Phys. Rev. Lett. {\bf 60}, 2058 (1988).

\bibitem{ni} X.Z. Ni and L.W. Bruch, Phys. Rev. B {\bf 33}, 4584 (1986). 

\bibitem{got} J.M. Gottlieb and L.W. Bruch, Phys. Rev. B {\bf 40}, 148 (1989). 

\bibitem{manou1} K. Nho and E. Manousakis, Phys. Rev. B {\bf 65}, 115409 (2002).

\bibitem{manou2} K. Nho and E. Manousakis, Phys. Rev. B {\bf 67}, 195411 (2003).

\bibitem{vives} E. Vives and P.A. Lindgard. Phys. Rev. B {\bf 47} 7431 (1993).

\bibitem{prl97} M.C. Gordillo and D.M. Ceperley, Phys. Rev. Lett. {\bf 79}, 
3010 (1997).   

\bibitem{bonin} M. Boninsegni, Phys. Rev. B {\bf 70}, 193411 (2004).  

\bibitem{deu3175} H. Freimuth, H. Wiechert, H.P. Schildberg, and H.J. Lauter
Phys. Rev. B. {\bf 42}, 587 (1990).   


\end{thebibliography}
\end{document}